\documentclass[a4paper]{article}

\usepackage{INTERSPEECH2019}

\usepackage{amssymb,amsmath,epsfig}
\usepackage{textcomp}
\usepackage{amsfonts,amsthm,amsopn}
\usepackage{url}
\usepackage[capitalise]{cleveref}
\usepackage{graphicx,caption,lipsum}
\usepackage{booktabs,multirow}
\usepackage{placeins}
\usepackage{algpseudocode,algorithm}
\usepackage{color}
\usepackage[table,xcdraw]{xcolor}
\usepackage{bbm}
\usepackage{enumerate}   

\usepackage[normalem]{ulem}

\ninept
\def\reg{{\rm\ooalign{\hfil
      \raise.07ex\hbox{\scriptsize R}\hfil\crcr\mathhexbox20D}}}

\title{Sound Event Detection in Multichannel Audio using Convolutional Time-Frequency-Channel Squeeze and Excitation}

\makeatletter
\def\name#1{\gdef\@name{#1\\}}
\makeatother 
\name{{\em } }

\name{{Wei Xia$^{1,2} $, Kazuhito Koishida
$^2$}\thanks{This work is done under Microsoft Research Internship program.}}
\email{wei.xia@utdallas.edu, kazukoi@microsoft.com}

\address{
$^1$ Center for Robust Speech Systems, University of Texas at Dallas, TX 75080 \\
$^2$ Microsoft Corporation, One Microsoft Way, Redmond, WA 98052
{\small \tt}}

\newcommand{\vct}[1]{\boldsymbol{\mathbf{#1}}} 

\graphicspath{{./}}

\begin{document}

\maketitle

\ninept

\begin{abstract}
In this study, we introduce a convolutional time-frequency-channel "Squeeze and Excitation" (tfc-SE) module to explicitly model inter-dependencies between the time-frequency domain and multiple channels.
The tfc-SE module consists of two parts: tf-SE block and c-SE block which are designed to provide attention on time-frequency and channel domain, respectively, for adaptively recalibrating the input feature map.
The proposed tfc-SE module, together with a popular Convolutional Recurrent Neural Network (CRNN) model, are evaluated on a multi-channel sound event detection task with overlapping audio sources: the training and test data are synthesized TUT Sound Events 2018 datasets, recorded with microphone arrays.
We show that the tfc-SE module can be incorporated into the CRNN model at a small additional computational cost and bring significant improvements on sound event detection accuracy. We also perform detailed ablation studies by analyzing various factors that may influence the performance of the SE blocks. We show that with the best tfc-SE block, error rate (ER) decreases from 0.2538 to 0.2026, relative 20.17\% reduction of ER, and 5.72\% improvement of F1 score. The results indicate that the learned acoustic embeddings with the tfc-SE module efficiently strengthen time-frequency and channel-wise feature representations to improve the discriminative performance.

\end{abstract}


\section{Introduction}
\label{sec:intro}



Sound event detection (SED) task involves labeling the time stamps of a sound event in audio streams and detecting the sound type. Speech and non-speech sounds such as laughter and music contains lots of useful information. Being able to detect environmental sound events in multi-channel audios can greatly help us understand surrounding acoustic environments and enables many applications such as audio surveillance and rare sound detection~\cite{kotus2014detection, foggia2016audio, crocco2016audio}.
It is also helpful to improve the performance of speech enhancement and separation systems if we could know the types of sounds~\cite{stowell2015detection,kong2018joint}. 
Robotic systems can employ SED for navigation and natural interaction with surrounding acoustic environments~\cite{takeda2016sound,he2018deep}. Smart home devices can benefit from it for environmental sound understanding~\cite{southern2017sounding, kao2018r}. Sound event detection is attracting more and more attention nowadays from these applications, as well as multimedia content retrieval~\cite{xu2008audio,jin2012event} and audio segmentation~\cite{kumar2016audio, tian2015use, wichern2010segmentation}.


Real life audio recordings typically have many overlapping sound events, the task of recognizing all the overlapping sounds is considered as polyphonic SED. Lots of efforts have been proposed to address this task by predicting frame-wise labels of each sound event class. Models like Gaussian Mixture Model (GMM)~\cite{atrey2006audio}, Hidden Markov Model (HMM)~\cite{mesaros2010acoustic}, Recurrent Neural Networks (RNN)~\cite{hayashi2017duration, parascandolo2016recurrent}, and Convolutional Neural Networks (CNN)~\cite{hershey2017cnn, zhang2015robust} have been explored extensively for this task. More recently successful results were obtained by stacking CNN, RNN and FC layers consecutively, referred jointly as the convolutional recurrent neural network (CRNN)~\cite{cakir2017convolutional}. 
In order to improve the recognition of overlapping
sound events, several multi-channel SED methods have also been proposed. For example, Sharath \cite{adavanne2017sound, adavanne2018sound, adavanne2018multichannel} showed that Deep Neural Network can directly learn from low-level features such as generalized cross-correlation with phase based weighting (GCC-PHAT) for multi-channel sound detection.

The basic building block for most SED models is the convolutional layer, which learns filters capturing local spatial patterns  along all the input channels and generates feature maps jointly encoding the time-frequency and channel information. 
A lot of work has aimed at improving the joint encoding of spatial and channel information~\cite{dai2017deformable}, but much less attention has been given towards encoding of time-frequency and channel-wise patterns independently with domain information. Some recent work attempted to address this issue by explicitly modeling the inter-dependencies between the channels of feature maps. One promising approach to accomplish this is a component called "Squeeze \& Excitation" (SE) block~\cite{hu2018squeeze, roy2018concurrent}, which can be seamlessly integrated into the CNN model. This SE block factors out the spatial dependency by global average pooling to learn a channel specific descriptor, which is used to rescale the input feature map to highlight only useful channels. 

In this study, we introduce a time-frequency-channel Squeeze and Excitation (tfc-SE) block for multi-channel sound event detection.
Different from the original SE block mentioned above, which uses global average pooling and only excites on the channel axis, we adapt and extend it to both the time-frequency and channel domains. 
Multi-channel audio contains different information at each time-frequency location, for example, we may pay more attention to high energy parts in the spectrogram.
We first introduce a time-frequency SE (tf-SE) block that aims to adaptively recalibrate the learned feature map. It does not change the receptive field, but provides time-frequency attention to certain regions. Further, we propose two methods to combine the channel wise SE and time-frequency SE. These methods aggregate the unique properties of each block and make feature maps to be more informative on both domains.
We show that with the best tfc-SE block, error rate of the SED system decreases from 0.2538 to 0.2026, relative 20.17\% reduction of ER. It also has 5.72\% relative improvement of F1 score compared to the original CRNN.

In the following sections, we describe the tfc-SE approach and corresponding baseline systems in \cref{sec:model}. We provide detailed explanations of our experiments in \cref{sec:exp}, as well as results and discussions in \cref{sec:result}. Finally we conclude our work in \cref{sec:conclusion}.

\begin{figure}[tbp]
    \centering
    \includegraphics[width=0.9\linewidth, height=5cm]{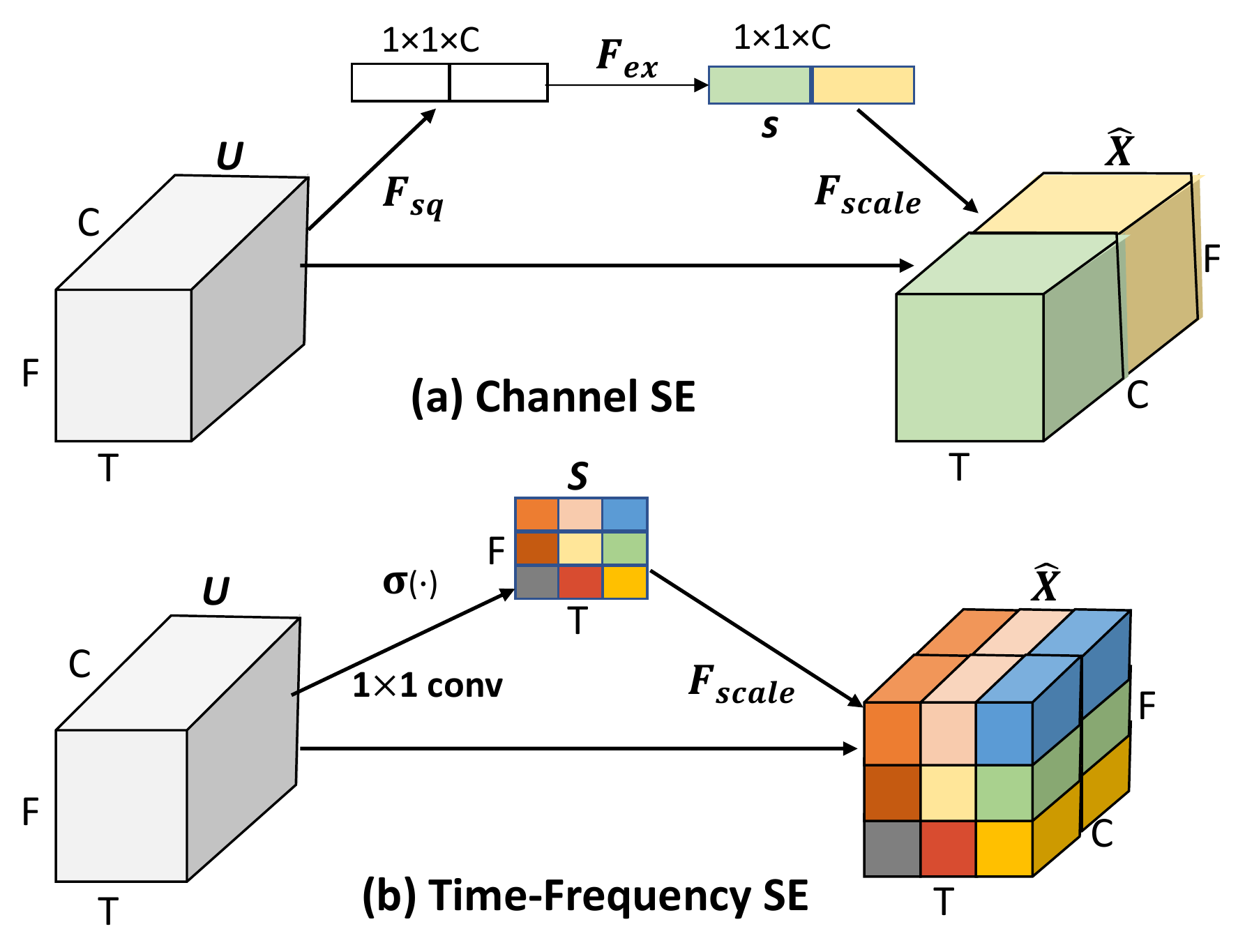}
    \caption{Channel SE and Time-Frequency SE blocks.}
    \label{fig:se1}
\end{figure}

\section{Sound Event Detection Systems}
\label{sec:model}

\subsection{Baseline Convolutional RNN}
We use a recently proposed Convolutional Recurrent Neural Network (CRNN)~\cite{cakir2017convolutional} to learn the acoustic representations of multi-channel audio signals for sound event detection. The CRNN model has three components, the convolutional layers to learn the time-frequency representations of audio waveform, the recurrent layers to learn the temporal information, and the final classification layer for sound event detection. The model configuration is given in \cref{tab:crnn}. We use three layers of 2D CNN to learn the shift invariant features from multi-channel spectorgrams. Each CNN layer has $P$ filters of a $3\times3$ kernel. After each CNN layer, we use a ReLU activation function and batch normalization to normalize the activations. A Max-pooling layer is applied on the frequency axis and the sequence length of $T$-frame input features are kept unchanged.

\begin{table}[hbp]
  \centering
  \caption{Baseline CRNN model architecture, when input are $T=256$ frames, 256 dimensional and 16 channels features, $P=64$ filters, $Q=128$ nodes, $R=128$ nodes, and $N=11$ sound types.}
    \begin{tabular}{lrr}
        \toprule

     Type  & \multicolumn{1}{l}{Filter shape}  & \multicolumn{1}{l}{Input shape}   \\
    \midrule

    Conv1 & $3\times 3\times16\times64$ & $256\times256\times16$  \\
    Maxpool1 & $1\times8$ &  $256\times256\times64$   \\
    Convol2 &  $3\times3\times16\times64$ & $256\times32\times64$  \\
    Maxpool2 & $1\times8$ &  $256\times4\times64$   \\
    Convol3 &  $3\times3\times16\times64$ &  $256\times4\times64$  \\
    Maxpool3 & $1\times2$ &  $256\times 2\times64$  \\
    Bi-GRU1 &  - &  $256\times128$  \\
    Bi-GRU2 &  - &  $256\times128$  \\
    FC1 &  $128\times128$ &  $256\times128$   \\
    FC2 &  $128\times11$ &  $256\times128$  \\
    \bottomrule
    \end{tabular}%
  \label{tab:crnn}%
\end{table}%

The output activation from CNN is further fed to bidirectional RNN layers which are designed to learn the temporal context information from the CNN output activations. Specifically, $Q$ nodes of Gated Recurrent Units (GRU) are used in each layer with tanh activations. For classification, we use two fully connected (FC) layers. The first FC layer has $R$ nodes each with linear activation. The last FC layer consists of $N$ nodes with sigmoid activations, each corresponding to one of the $N$ sound event classes to be detected.

In the following sections, we introduce the "Squeeze \& Excitation" (SE) blocks. We insert the SE block after each convolutional layer to adaptively recalibrate the feature representations.

\begin{figure}[tbp]
    \centering
    \includegraphics[width=0.9\linewidth, height=5cm]{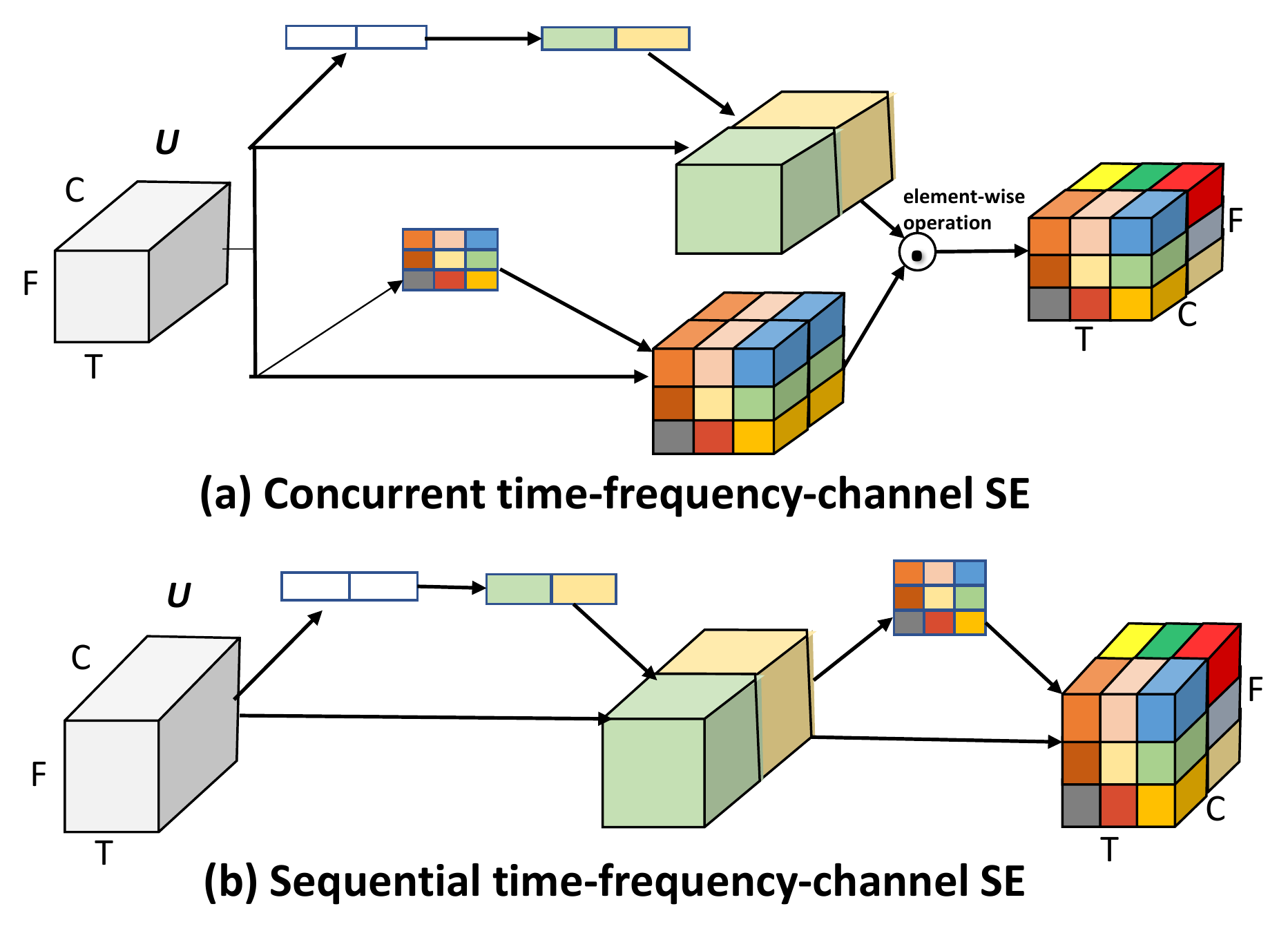}
    \caption{Concurrent time-frequency-channel SE and sequential time-frequency-channel SE blocks.}
    \label{fig:se2}
\end{figure}

\subsection{Channel wise squeeze and excitation}
\subsubsection{Squeeze operation}
The channel-wise SE block, c-SE, is illustrated in \cref{fig:se1} (a).
In order to model the inter-dependencies between multiple channels of audio signals, we firstly define a squeeze operation to embed the global time-frequency information into a channel descriptor.
We consider the input feature map of the multi-channel audio as $\mat U=[\mat U_1, \mat U_2, ..., \mat U_c]$ where $\mat U_c\in \mathbb{R}^{T\times F}$ is the feature matrix of channel $c$.
We use a global average pooling layer to generate a channel-wise vector $\vct z\in \mathbb{R}^{1\times 1\times C}$ with its $c$-th element, 
\begin{align}
z_c = \mathcal{F}_{sq}(\mat U_c) = \frac{1}{T\times F}\Sigma_{i}^{T}\Sigma_{j}^{F}\mat U_c(i,j) 
\end{align}
This operation embeds the global time-frequency information into the vector $\vct z$. This vector contains statistics that are expressive of the whole time-frequency input.

\subsubsection{Excitation operation}
\label{sec:excitation}
In order to capture channel wise dependencies, a gating mechanism with a sigmoid activation function is used to learn a non-linear relationship between channels,
\begin{align}
\vct s = \mathcal{F}_{ex}(\vct z, \mat W) = \sigma (g(\vct z, \mat W)) = \sigma(\mat W_2 \delta(\mat W_1 \vct z))
\end{align}
where $\mat W_1 \in \mathbb{R}^{\frac{C}{r}\times C}$ and $\mat W_2\in \mathbb{R}^{C\times \frac{C}{r}}$ are weights of two fully-connected layers and $\delta$ is the ReLU operator. The dimensional reduction factor $r$ indicates the bottleneck in the channel excitation block. 
Note that the original channel dimension is recovered by the second FC layer.
With a sigmoid layer $ \sigma$, the channel-wise attention vector $\vct s$ is obtained. Finally we recalibrate $\mat U$ with the attention vector as,
\begin{align}
\hat{\mat X}_c = \mathcal{F}_{scale} (\mat U_c, s_c) = s_c\cdot \mat U_c
\end{align}
 $\hat{\mat X}=[\hat{\mat X}_1, \hat{\mat X}_2, ..., \hat{\mat X}_C]$ is the final channel wise recalibrated features.
In this block, the input features are attentively scaled so that
important channels are emphasized and less important one are diminished.

\subsection{Time-frequency squeeze and excitation}

Here we introduce the time-frequency wise squeeze and excitation block (tf-SE), shown in \cref{fig:se1} (b). The concept is similar to the c-SE and the main difference is to compress the feature map $\mat U$ along the channel axis using a 1-by-1 convolution~\cite{lin2013network}. The excitation operation is performed on the time-frequency map. We assume that the time frequency space may contain more information for SED.


We write the input feature map $\mat U$ in an alternate form,
\begin{align}
\mat U =
\begin{bmatrix}
\vct u^{1,1} & \vct u^{1,2} & ... & \vct u^{1,F} \\
\vct u^{2,1} & \vct u^{2,2} & ... & \vct u^{2,F} \\
\vdots & \vdots & \ddots & \vdots \\
\vct u^{T,1} & \vct u^{T,2} & ... & \vct u^{T,F}
\end{bmatrix}
\end{align}
where $\vct u^{i,j}\in \mathbb{R}^{1\times 1\times C}$ is the feature bin at the time-frequency location $(i,j)$. The squeeze operation of the tf-SE is done using a 1-by-1 convolution which can be represented by the linear combination of all channels $C$ at a location $(i,j)$, 
\begin{align}
s^{i,j} = \sigma((\vct w^{i,j})^T \cdot \vct u^{i,j})
\end{align}
where $\vct w^{i,j}$ is the filter coefficients and $s^{i,j}$ is the $(i,j)$-th element of the squeezed matrix $\mat S\in \mathbb{R}^{T\times F}$. 
We use a sigmoid function $\sigma$ to limit the range of the matrix $\mat S$ to [0, 1], which is used to recalibrate the features on the time-frequency domain. Each value $s^{i,j}$ corresponds to the relative importance in the time-frequency space.

Similar to the c-SE, the excitation of the tf-SE is carried out as
\begin{align}
\hat{\vct x}^{i,j} = \mathcal{F}_{scale} (\vct u^{i,j}, s^{i,j})
\end{align}
and $\hat{\vct x}^{i,j}$ is the $(i,j)$-th element of the recalibrated output $\hat{X}$.

\subsection{Concurrent and sequential time-frequency-channel SE}

Each of the above explained c-SE and tf-SE blocks has its unique properties. The c-SE blocks recalibrates the channel information by incorporating global time-frequency information.
On the other hand, the tf-SE block generates an time-frequency attention map, indicating where the network should focus more to aid the sound classification.

We propose two ways to combine the complementary information from these two SE blocks to form the time-frequency-channel SE (tfc-SE):
1) Concurrent recalibration, illustrated in \cref{fig:se2} (a) and 2)
Sequential recalibration in order of channel first and then time-frequency as in \cref{fig:se2} (b).

For the concurrent tfc-SE, we present four ways to aggregate the c-SE and the tf-SE blocks.
\begin{enumerate}[(i)]
\itemsep0em 
\item  Addition: the two recalibrated feature maps are added element wise with equal weights.
\item Multiplication: the feature maps are multiplied element wise. 

\item Maximization: each location $(i,j,c)$ of the output feature map has the maximum activation of the two feature maps element-wise.
\item  Concatenation: 
the two feature maps are concatenated along the channel axis, which means that the volume of the input feature is twice larger.
\end{enumerate}
The concurrent tfc-SE and sequential tfc-SE with these aggregation strategies will be evaluated in the following sections.


\section{Experimental Setup}
\label{sec:exp}

\subsection{Dataset}
To study the effectiveness of the CRNN model with the tfc-SE module in multi-channel sound event detection, we used the synthetic eight-channel TUT Sound Events 2018 - Circular array, Reverberant and Synthetic Impulse Response (CRESIM) dataset~\cite{adavanne2018sound}. The dataset synthesizes the DCASE 2016 task 2 dataset~\cite{dcase2016}, which has 11 isolated sound event classes such as speech, door slam, phone ringing, coughing, and keyboard.

Each of the sound classes has 20 examples, of which 16 are randomly chosen for the training set and the rest 4 for the test set, in total 176 examples from 11 classes for training, and 44 for testing. We selected the O3 subset which has maximum three temporally overlapping sources.
It consists of three cross-validation splits with 240 training and 60 testing  recordings of length 30 sec sampled at 44100 Hz. 
In this dataset, circular microphone array recording is simulated with 8 omnidirectional microphones equally spaced on 5cm radius. More details can be found in~\cite{adavanne2018sound}.

\subsection{Feature extraction}
On frame basis, we compute the magnitude and phase of spectrograms and concatenate them along the channel axis as input features. A Hamming window of length $M$ with 50\% overlap is used to extract the spectrogram from each audio channel.
The zeroth bin is excluded from the spectrogram, so that each frame produces a $M/2 \times 16$ feature matrix. The input feature is mean and variance normalized on the frame level. 
A sequence of the $T$-frame spectrograms are stacked and fed into the network.

\subsection{Evaluation metrics}



 We evaluated our sound detection model using the standard polyphonic SED metrics, error rate (ER) and F-score calculated on segments of one second with no overlap as proposed in \cite{mesaros2016metrics}. F1 (ideally F1 = 1)  is based on true and false positives, and the error rate (ER) (ideally
ER = 0) is based on the total number of active sound event
classes in the ground truth. A joint SED error score can be considered as $S_{SED}= (ER + 1-F)/2$.

\subsection{Model configuration and training}

We explored various CRNN models and feature configurations. It was found from preliminary experiments that the best input feature setting is a 
window length of $M=512$ and sequence length of $T=256$ (equivalent to 1.486 sec).
The CRNN model was trained on the CRESIM dataset for 1000 epochs with a batch size of 64.
Early stopping was applied if there is no improvement on the $S_{SED}$ score after 100 epochs.
We built a various models with different parameters such as the number of nodes of CNN/RNN/FC and CNN filter size.
The best baseline setting can be found in \cref{tab:crnn}.
The last fully connected layer has 11 nodes with sigmoid activation and the cross entropy loss for classification was employed for network training. 
We used an Adam optimizer with the learning rate $10^{-3}$. 
For each sound event class all the predicted probabilities in the sequence ($T=256$) were examined at the frame level and the class detection was claimed as true if there is any probability which is larger than a threshold of 0.5.



\section{Results and Discussions}
\label{sec:result}

\subsection{Experimental results}
In order to thoroughly evaluate our proposed methods, we will conduct detailed ablation analysis in this section. We first perform experiments on the c-SE, followed by the tf-SE, and then the sequential tfc-SE for sound event detection. We further investigate the aggregation strategies for the concurrent tfc-SE. 
Finally we analyze various factors that may affect the performance of SE blocks.

From \cref{tab:sed_results}, we observe that our proposed channel wise SE (c-SE) block improves both F1 and ER compared with the original CRNN model. After utilizing the c-SE block, overall error score $S_{SED}$ of the model decreases from 0.2285 to 0.2194, relative 3.98\% gain. This approach also achieves relative 6.30\% improvement in ER. We also find a consistent performance improvement with the time-frequency SE (tf-SE) block, relative 4.08\% and 12.02\% improvement in F1 and ER, respectively, compared with the original CRNN model. 
The above results indicate that the tf-SE is more efficient than c-SE, which aligns with our assumption that the time-frequency space may have more meaningful information than channel for the SED task.

Finally, we test the tfc-SE block which combines the c-SE and t-SE activations. Both the concurrent and sequential models improve the performance of the original CRNN model by a large margin. The best result is obtained with the sequential tfc-SE block, which achieves 84.23\% F1 score and 0.2026 error rate, relative 5.72\% and 20.17\% improvement compared with the original CRNN. In terms of the overall $S_{SED}$ score, the sequential tfc-SE block outperforms the original CRNN model by 21.18\% relatively.

\vspace{-2ex}

\begin{table}[htbp]
  \centering
   \caption{Averaged SED results on CRESIM overlap 3 three data splits using the CRNN model with time-frequency-channel squeeze and excitation blocks.}
    \begin{tabular}{l|rrrr}
        \toprule

       Model   & \multicolumn{1}{l}{F1 (\%)} & \multicolumn{1}{l}{ER}  & \multicolumn{1}{l}{$S_{SED}$} \\
    \midrule

    CRNN & 79.67 & 0.2538 & 0.2285 \\
    +c-SE & 79.91 & 0.2378 & 0.2194 \\
    +tf-SE &  82.95 &  0.2233 & 0.1952  \\
    +tfc-SE concurrent & 83.57 &  \textbf{0.1982} & 0.1812 \\
    +tfc-SE sequential & \textbf{84.23} & 0.2026 & \textbf{0.1801} \\
    \bottomrule

    \end{tabular}%
  \label{tab:sed_results}%
\end{table}%

\vspace{-4ex}
\subsection{Aggregation strategies}

\begin{table}[bp]
  \centering
  \vspace{-3ex}
    \caption{SED results on CRESIM overlap 3 split 1 subset using  CRNN + tfc-SE concurrent block with different aggregration operations.}
    \begin{tabular}{l|rrrr}
    \toprule

     Aggregation     & \multicolumn{1}{l}{F1(\%)} & \multicolumn{1}{l}{ER}  & \multicolumn{1}{l}{$S_{SED}$} \\
    \midrule
    Addition & 84.92 & 0.1791 & 0.1649 \\
    Multiplication & 84.48 & 0.1959  & 0.1756 \\
    Maximization & \textbf{85.79}  & \textbf{0.1703}  & \textbf{0.1562} \\
    Concatenation & 85.26  & 0.1841  & 0.1657 \\

    \bottomrule

    \end{tabular}%
  \label{tab:agg}%
\end{table}%

We further investigate the aggregation strategy of the concurrent tfc-SE block, among the four choices. We observe from \cref{tab:agg}  that all aggregation methods increase the SED performance against the original CRNN. Using maximization provides the best performance. As concatenation aggregation increases the model complexity by the doubled number of channels, maximization operator looks a better choice for a lower computational cost. For all other experiments, we use the maximization-based aggregation for tfc-SE blocks.

\vspace{-1ex}
\subsection{Sensitivity to dimension reduction ratio}
The reduction ratio $r$ introduced in \cref{sec:excitation} is an important hyperparameter that allows us to vary the capacity and computational cost of the c-SE blocks in the model. \cref{tab:ratio} reveals that the performance does not improve monotonically with increased capacity. This probably comes from overfitting due to a larger model capacity.
We find that $r=8$ is a good option and we use it for our other experiments. 


\begin{table}[htbp]
  \centering
  \vspace{-1ex}
    \caption{SED results on CRESIM overlap 3 split 1 subset using CRNN + tfc-SE concurrent block with different dimension reduction ratios.}
    \begin{tabular}{l|rrrrr}
        \toprule

         Ratio r & \multicolumn{1}{l}{F1(\%)} & \multicolumn{1}{l}{ER}   & \multicolumn{1}{l}{$S_{SED}$}  \\
    \midrule

    2 & 85.07 & 0.1751 & 0.1622  \\
    4 & 85.59 & 0.1881 & 0.1661 \\
    8 & \textbf{85.79}  & \textbf{0.1703}  & \textbf{0.1562} \\
    16 & 83.51 & 0.1957  &  0.1803 \\

    \bottomrule
    \end{tabular}%
 
  \label{tab:ratio}%
\end{table}%

\vspace{-4ex}
\subsection{Squeeze and excitation operator}
For the squeeze operator, we examine the significance of using global average pooling as opposed to global max pooling (while keeping the excitation operator sigmoid) in \cref{tab:squeeze}. Though both max and average pooling are effective, average pooling achieves a little better result and we use it all in our paper. We next assess the excitation operator. Two  options, ReLU and tanh, are experimented by replacing from the sigmoid (with leaving the squeeze operator global average pooling). It suggests that it is important to choose the sigmoid operator in order to make the tfc-SE block effective.

\begin{table}[htbp]
  \centering
  \vspace{-1ex}
    \caption{SED results on CRESIM overlap 3 split 1 subset using CRNN + tfc-SE concurrent block with different squeeze and excitation operators.}
    \begin{tabular}{l|rrrr}
    \toprule
     Operator & \multicolumn{1}{l}{F1(\%)} & \multicolumn{1}{l}{ER} & \multicolumn{1}{l}{$S_{SED}$} \\
    \midrule
    Max & 82.89 & 0.1971  & 0.1841  \\
    Avg & 85.79  & 0.1703  & 0.1562 \\
    \midrule
    \midrule
    ReLU & 83.42 & 0.1799 & 0.1728  \\
    Tanh & 82.58  & 0.2196  & 0.1969 \\
    Sigmoid & 85.79  & 0.1703  & 0.1562 \\

    \bottomrule

    \end{tabular}%

  \label{tab:squeeze}%
\end{table}%

\vspace{-4ex}
\subsection{Model complexity}
The original CRNN model has 496,587 parameters.   The tfc-SE  block  requires  only  0.7\%  additional  parameters,  which  is 500,067 in total.

\vspace{-2ex}
\section{Conclusions}
\label{sec:conclusion}

In this paper, we proposed a convolutional time-frequency-channel squeeze and excitation block for multi-channel sound event detection, in order to model the feature inter-dependencies between channels and the time-frequency locations. The tfc-SE block was inserted after each convolution layer of the CRNN model. The proposed method was evaluated on the CRESIM dataset and it was shown to improve the original CRNN model in terms of both F1 score and error rate by a large margin. These results indicated that the tfc-SE block effectively recalibrates the feature maps adaptively by emphasizing more important channels and time-frequency locations.

\vfill\pagebreak

\newpage

\bibliographystyle{IEEEtran}
\bibliography{is2019_sed.bib}

\end{document}